# Detailed and simplified plasma models in combined-cycle magnetohydrodynamic power systems

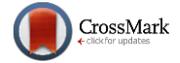

Osama A. Marzouk *

*College of Engineering, University of Buraimi, Al Buraimi, Oman*

A B S T R A C T

Magnetohydrodynamics (MHD) is a subject concerned with the dynamics of electrically conducting fluids (plasma) and can be applied in electric power generation. As a unique technology for producing direct-current electricity without moving parts, it can be utilized within a high-temperature topping power cycle to be combined with a traditional bottoming power cycle, forming a combined-cycle MHD system. This study presents governing equations for the electric field and current density field within a moving plasma subject to an applied magnetic field. The modeling equations are described at four descending levels of complexity. Starting with the first level of modeling, which is the most general case, where no assumptions are made regarding the electric field, plasma velocity field, applied magnetic field, or electrode geometry. In the second level of modeling, the magnetic field is treated as one-dimensional. In the third level of modeling, a specific Faraday-type magnetohydrodynamics plasma generator channel is considered, having two continuous electrodes acting as parallel constant-voltage terminals. In the fourth (and simplest) level of modeling, an additional approximation is made by setting the Hall parameter to zero and replacing all vector fields with scalar quantities. For that simplest model, a representative set of operation conditions (electric conductivity 20 S/m, temperature 2800 K, supersonic plasma gas speed 2000 m/s with Mach 2.134, and magnetic flux density 5 T) shows that the optimum idealized electric power that can be extracted from a unit volume of plasma is estimated as 500 MW/m$^3$. This is a much larger volumetric power density than typical values encountered in reciprocating piston-type engines (0.2 MW/m$^3$) or rotary gas turbine engines (0.5 MW/m$^3$). Such an extremely high power density enables very compact power generation units.

## 1. Introduction

Global population growth and technological development led to an increasing demand for electricity as a critical utility for society (Sen et al., 2021). Large-scale commercial production of electricity can utilize one of several primary energy sources, with a power plant converting this natural energy source into electricity. Many well-established technologies for electricity generation are based on Faraday's law of electromagnetic induction, where an electromotive force (developed voltage) is induced in a coil if it is subject to a changing magnetic field relative to it (Smit et al., 2018). A key element of these electricity generation technologies is an alternating-current (AC) mechanical turbogenerator, where a number of coils are made to rotate (using input mechanical rotation shaft power) within an applied magnetic field (Danilevich et al., 2010). Steam power plants, diesel power plants, gas turbine power plants, hydroelectric (hydropower) power plants, and nuclear power plants utilize the conventional method of mechanical turbogenerators (Raja and Srivastava, 2006). Other non-conventional electricity generation technologies include thermoelectric generators (Liu et al., 2023), fuel cells (Aminudin, 2023), photovoltaic solar cells (Marzouk, 2022), and magnetohydrodynamics (MHD) generators (Woodside et al., 2012; Marzouk, 2018a).

Conventional turbogenerator-based electricity generation technologies are currently the dominant type of electricity generation in the global power sector and are expected to remain as such in the near future. However, non-conventional methods of electricity generation may become gradually more exploited given some advantages they offer, such as reduced harmful emissions, modularity, or portability (Zhan et al., 2023).

* Corresponding Author.
Email Address: osama.m@uob.edu.om
https://doi.org/10.21833/ijaas.2023.11.013
 Corresponding author's ORCID profile:
https://orcid.org/0000-0002-1435-5318

The current study focuses on plasma-based MHD power plants as a concept for generating direct-current (DC) electricity without moving parts. In the MHD generator, the conductor that is subject to a magnetic field is not a solid coil, but a gaseous medium consisting of an electrically conducting gas (plasma). The electric conductivity of a neutral gas can be achieved by seeding an alkali metal vapor, such as cesium (Cs, atomic number 55) or potassium (K, atomic number 19) with a small fraction, such as only 1% by mole. These metal elements have relatively low ionization energies, such as 3.894 eV for cesium, which is the element having the smallest ionization energy, and 4.341 eV for potassium, which is the element having the fourth smallest ionization energy (NCBI, 2023). The ionization of the seeded metal as well as any minor ionization of the original gases in the plasma occurs at elevated temperatures (Zhou et al., 2023). Alternatively, an electrically conducting liquid metal may be used as a working medium, and in that case seeding is not necessary (Cheng et al., 2022). There are different proposed designs for MHD power generation, such as the open loop or linear channel category (Bera et al., 2022) versus the closed-loop or disc category (Okamura et al., 1985; Okunoa et al., 1999).

The idea of MHD plasma power generation is not new and has been studied experimentally, analytically, or computationally for years (Blackman et al., 1961; Wright, 1963; Jones, 1985; Rosa et al., 1991; Borghi et al., 1992; Davidson, 2001; Panchenko, 2002; Kayukawa, 2004; Ishikwa et al., 2007; Sarkar, 2017). Despite this, the concept of plasma power plant has not been used practically at commercial levels because economically it is not favorable compared to other technologies of power plants that have comparable efficiency but with less initial and operational cost, such as combined cycle gas turbines (CCGT) power plants, whose LHV (lower heating value) efficiency can reach nearly 50% (Takeishi and Krewinkel, 2023). However, plasma generators still have advantages such as the lack of moving parts, and the potential of reducing particulate matter (PM), sulfur dioxide ($SO_2$), and nitrogen oxides (NOx) compared to a coal power plant with air-firing.

Similar to turbogenerators, which can operate reversely as electric motors; plasma generators can also operate in a reverse mode as magnetoplasmadynamic (MPD) thrusters or MHD thrusters for propulsion. In this case, the electrically conducting fluid is accelerated under the influence of an applied magnetic field as well as an applied electric field, due to the Lorentz force effect (Cébron et al., 2017; Chen et al., 2022; Zhao et al., 2022; Everett and Ryan, 2023).

Because MHD channels operate at very high temperatures with a peak temperature near 3000 K, their exhaust gases can be as hot as an air-fuel flame, with temperatures around 2300 K. Therefore, it is not feasible to operate these generator channels as a standalone power plant because of the excessive waste heat released in the exhaust gas. Instead, such channels should be used within an "MHD system", which is a combined-cycle power plant with two related power cycles. The higher-temperature cycle (or the topping cycle) is the plasma generator. Its hot exhaust gas is used as an input heat source to a lower-temperature cycle (or the bottoming cycle), which can be a conventional steam power plant with turbogenerators driven by steam turbines according to a Rankine power cycle (Marzouk, 2023), or a conventional gas-fired power plant with turbogenerators driven by gas turbines according to a Brayton power cycle (Esmaeilzadehazimi et al., 2019).

This paper considers the analysis of the working of the plasma generator (MHD generator) as a DC power plant. This analysis is conducted under 4 levels of approximations. First, the three-dimensional equations are provided, where the plasma velocity vector, the current density vector, and the electric field vector are three-dimensional. Second, a reduced version is presented when the applied external magnetic field is one-dimensional (perpendicular to the longitudinal direction of the bulk motion of the plasma). Third, the electric current density is assumed to be two-dimensional, and the electric field is assumed to be one-dimensional. This corresponds to a particular design of plasma generators, which is the linear rectangular Faraday type (having two continuous electrodes at opposite sides, one cathode and one anode) and a rectangular channel (having a uniform distance between the two electrodes). In addition, a constant voltage is assigned at each electrode. When viewed from the side, the channel would have a rectangular layout with the two electrodes forming the upper and lower edges. This specific design has the feature that the electrodes are parallel, and thus the electric field vectors within the plasma are parallel also, but perpendicular to the electrodes. The longitudinal direction (along the channel) is the direction of the bulk motion of the plasma (the ionized gas), which when flows perpendicular to the applied magnetic field, an electric field is induced. A resultant electric current density occurs within the plasma, which can be collected at the electrodes to supply electric current to an external electric load. Fourth, the simplest version of the analysis is derived for the case where the current density vector is also treated as a one-dimensional quantity. Thus, all involved vector fields in the MHD problem become one-dimensional, which enables writing an analytical expression for the maximum electric power output per unit volume of the channel, at recommended operating conditions. This is viewed as a theoretical upper limit for the multi-dimensional configurations. The present work is limited to the plasma electromagnetic behavior within the MHD channel. The electric characteristics of the plasma are no longer relevant after leaving the channel as an exhaust gas.

The usefulness of the present paper can be manifested through the explanation of the uncommon concept of combined-cycle MHD systems, as well as the use of original sketches to explain the operation of MHD channels for the purpose of electricity generation. The careful arrangement of the plasma models covered here allows the reader to understand involved assumptions at each level, as well as to recognize how a particular model is related to a higher (more complex) one or to a lower (more simplified) one. A quantitative example is also provided to quantify the potential superiority of MHD channels as sources of electric power.

**2. MHD system**

Fig. 1 illustrates a typical MHD system. Combustion takes place in the topping cycle only. The oxidizer can be either air in the case of conventional air-combustion, or oxygen separated from air through an air separation unit (ASU) for higher-temperature oxy-combustion. The seed material added to the plasma before it enters the MHD channel is recovered after the plasma exits the channel so that this material can be reused. The topping cycle (plasma generator) has 5 or 6 main components, as follows:

1. Compressor (to increase the pressure and density of the oxidizer)
2. Preheater for the oxidizer (optional component, useful in the case of air-combustion to increase the plasma temperature, but not necessary for oxy-combustion because the flame and combustion products already have an elevated temperature compared to air-combustion)
3. Combustor
4. Supersonic nozzle (such that the produced plasma gas is accelerated to a supersonic speed, exceeding the local speed of sound)
5. MHD channel
6. Supersonic diffuser (to decelerate the supersonic plasma leaving the supersonic MHD channel, from a higher speed exceeding the speed of sound to a lower speed below the speed of sound).

If the bottoming cycle represents a steam power plant, it consists of 5 main components, as follows:

1. Heat recovery steam generator (HRSG), which utilizes part of the heat in the hot plasma gas leaving the topping cycle to convert water into hot steam
2. Water pump
3. Steam turbine
4. Condenser (which converts the vapor steam into liquid water for reuse in a closed water loop)
5. Turbogenerator (driven by the steam turbine)

The electric power output from the plasma generator is in a direct-current (DC) form, with a fixed positive and negative polarity. This requires an auxiliary inverter to convert it into a sinusoidal alternating-current (AC) form. Another AC electric output comes from the bottoming cycle. The turbogenerator of the bottoming cycle does not need an inverter because it is able to naturally produce an AC electric waveform.

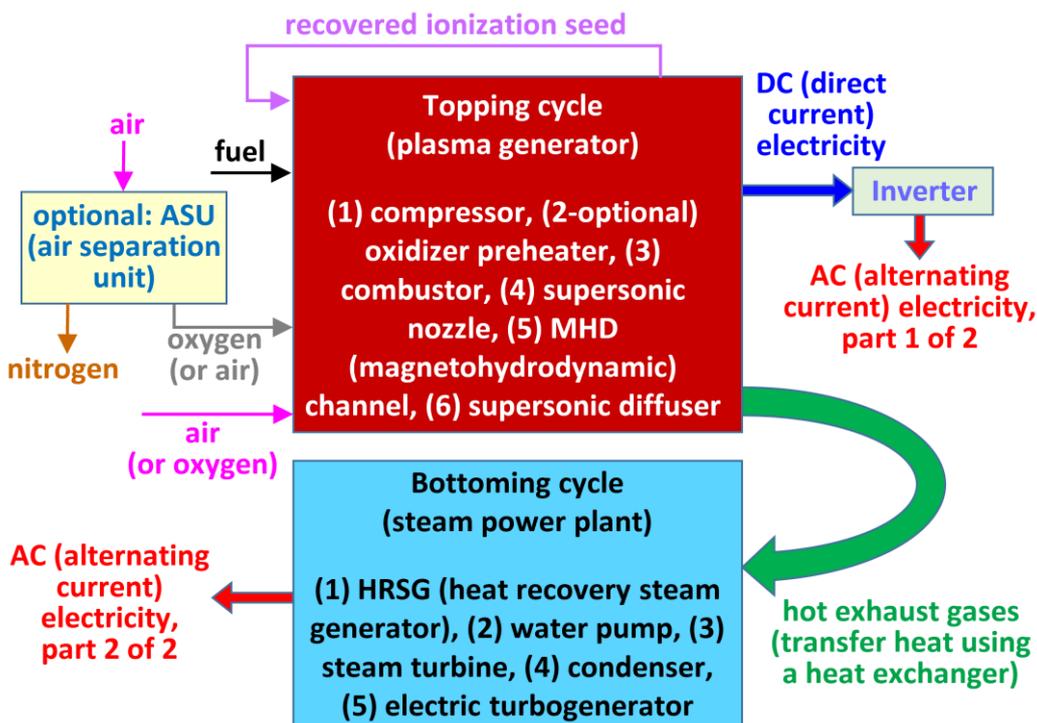

**Fig. 1:** Components of a typical MHD system with a bottoming steam power cycle

## 3. Plasma channel

Fig. 2 illustrates the elements of a typical rectangular MHD channel. The axes are chosen such that the x-axis is aligned with the bulk velocity of the plasma gas along the channel, while the y-axis is perpendicular to the surfaces of the two parallel electrode plates, and the z-axis is the direction of the applied external magnetic field (pointing from a north magnetic pole to a south magnetic pole).

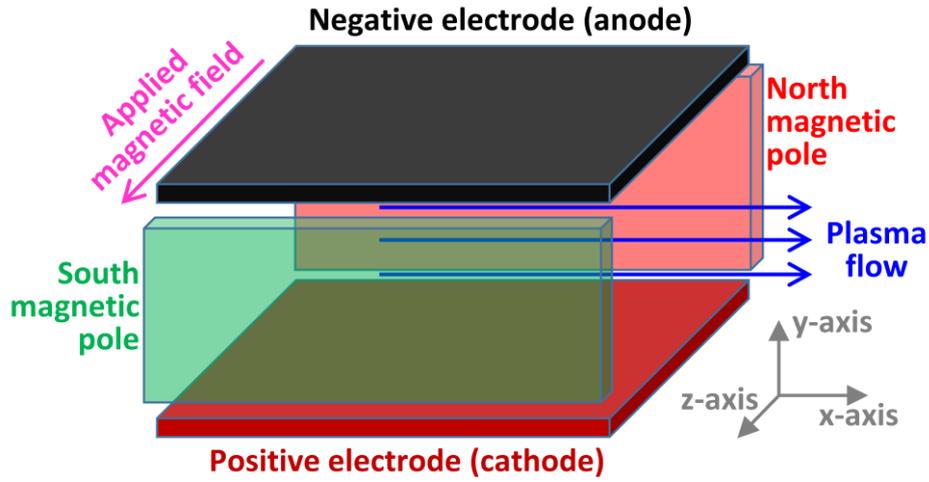

**Fig. 2:** Geometry of a rectangular MHD channel

## 4. General three-dimensional plasma equations

### 4.1. Generalized Ohm's law with three-dimensional magnetic field

In this subsection, all vector fields are treated as three-dimensional vectors, even the magnetic flux density vector field $\vec{B}$ (Messerle, 1995). This leads to the most general form of plasma electromagnetic equations These are considered a starting point, from which reduced simpler versions are to be derived for specific cases later in the current study.

The resultant electric field seen by the plasma (hence the subscript "P") is:

$$\vec{E_P} = \vec{E_L} + \vec{E_S} \tag{1}$$

where, $\vec{E_L}$ is an electric field vector caused by the difference in voltage of the two electrodes that are possibly connected to a powered external electric load (hence the subscript "L") and the induced electric field $\vec{E_S}$ acts as an electric field source (hence the subscript "S"), analogous to a voltage-source battery in traditional DC circuits (but here we speak of an electric field vector, rather than a scalar voltage quantity). This produced electric field vector is established due to the electromagnetic induction effect for the moving plasma gas as it is influenced by the applied magnetic flux density field $\vec{B}$. This induced electric field is related to the plasma velocity vector $\vec{u}$ and the magnetic flux density vector $\vec{B}$ as

$$\vec{E_S} = \vec{u} \times \vec{B} \tag{2}$$

substituting Eq. 2 into Eq. 1 gives:

$$\vec{E_P} = \vec{E_L} + \vec{u} \times \vec{B} \tag{3}$$

An electron drift velocity vector is caused by the electron mobility of the electrons as they are influenced by the above plasma-seen electric field as well as by the applied magnetic field. This electron drift velocity vector is:

$$\vec{d} = -\mu \left(\vec{E_P} + \vec{d} \times \vec{B}\right) \tag{4}$$

In the above equation, the minus sign is inserted because the electrons have a negative charge. Theoretically, there is also another drift for ions, but this is reasonably neglected because of the much larger mass of ions compared to electrons, making ions relatively immobile. Multiplying the above equation by (–n e) gives:

$$-n\,e\,\vec{d} = n\,e\,\mu \left(\vec{E_P} + \vec{d} \times \vec{B}\right). \tag{5}$$

But

$$-n\,e\,\vec{d} = \vec{j} \tag{6}$$

and

$$n\,e\,\mu = \sigma. \tag{7}$$

Thus, using Eqs. 6 and 7 into Eq. 5 gives:

$$\vec{j} = \sigma \left(\vec{E_P} + \vec{d} \times \vec{B}\right). \tag{8}$$

Expanding the brackets gives:

$$\vec{j} = \sigma \vec{E_P} + \sigma \vec{d} \times \vec{B}. \tag{9}$$

Using Eq. 7, the second term on the right-hand side of the above equation can be rewritten as:

$$\sigma \vec{d} \times \vec{B} = \mu\,n\,e\,\vec{d} \times \vec{B}. \tag{10}$$

Using Eq. 6, the above equation can be rewritten such that the current density vector $\vec{J}$ appears instead of electron drift velocity vector $\vec{d}$. Thus:

$$\sigma \vec{d} \times \vec{B} = \mu\, n\, e\, \frac{\vec{J}}{-n\,e} \times \vec{B}. \tag{11}$$

This can be simplified to:

$$\sigma \vec{d} \times \vec{B} = -\mu \vec{J} \times \vec{B}. \tag{12}$$

Therefore, the above Eq. allows rewriting Eq. 9 as:

$$\vec{J} = \sigma \vec{E_P} - \mu \vec{J} \times \vec{B} \tag{13}$$

The above vectorial equation is a generalized (or extended) version of Ohm's law, applicable to a three-dimensional conductor media, in contrast with the classical scalar Ohm's law that is related to solid one-dimensional conducting wires. The particular form of the generalized Ohm's law in Eq. 13 utilizes the mobility in the second term of its right-hand side.

The electron mobility and the magnitude of the magnetic flux density are related through the Hall parameter β as:

$$\beta = \mu\, B. \tag{14}$$

Therefore

$$\mu = \frac{\beta}{B}. \tag{15}$$

Therefore, Eq. 13 can be rewritten as:

$$\vec{J} = \sigma \vec{E_P} - \beta \vec{J} \times \frac{\vec{B}}{B}. \tag{16}$$

The quantity $\vec{B}/B$ is a unit vector in the direction of the magnetic flux density field. Eq. 16 is another form of the generalized Ohm's law, which utilizes the Hall parameter (instead of the electron mobility) in the second term of its right-hand side.

### 4.2. Generalized Ohm's law with one-dimensional magnetic field

In this subsection and in the rest of the analysis covered in this paper, a modeling simplification is made, which is treating the magnetic flux density field $\vec{B}$ as a one-dimensional vector, pointing in the positive z-axis. It should be noted that this magnetic field is not influenced by the plasma, but is independently controllable. Thus, this simplification is acceptable, and having a magnetic field with parallel magnetic field lines in one direction is possible (Markoulakis et al., 2022).

With this simplification, the quantity $\vec{B}/B$ becomes a unit vector in the positive z-axis, which is denoted here by $\hat{k}$. Thus, Eq. 16 can be simplified to:

$$\vec{J} = \sigma \vec{E_P} - \beta \vec{J} \times \hat{k}. \tag{17}$$

The above expression is an implicit equation in the electric current density vector $\vec{J}$, which appears on both the left-hand side and the right-hand side.

For a general three-dimensional electric current density vector $\vec{J}: (J_x, J_y, J_z)$, and a general three-dimensional plasma-seen electric field vector $\vec{E_P}: (E_x, E_y, E_z)$, the 3 scalar components of the vector Eq. 17 are:

$$J_x = \sigma E_x - \beta J_y \tag{18a}$$
$$J_y = \sigma E_y + \beta J_x \tag{18b}$$
$$J_z = \sigma E_z \tag{18c}$$

Eq. 18c is in an explicit uncoupled form with respect to $J_z$, while Eqs. 18a and 18b represent a coupled implicit system of equations in $J_x$ and $J_y$. Through mathematical manipulation, it is possible to rewrite Eqs. 18a and 18b in an explicit uncoupled form for $J_x$ and $J_y$, where each component of the electric current density has an explicit dependence on the x-component and the y-component of the electric field vector, as well as on the Hall parameter and on the electric conductivity of the plasma. The uncoupled version is:

$$J_x = \frac{\sigma}{1+\beta^2}(E_x - \beta E_y) \tag{19a}$$
$$J_y = \frac{\sigma}{1+\beta^2}(E_y + \beta E_x). \tag{19b}$$

For a general three-dimensional load-electrodes electric field vector $\vec{E_L}: (E_{Lx}, E_{Ly}, E_{Lz})$, and a general three-dimensional plasma gas velocity vector $\vec{u}: (u, v, w)$, the 3 scalar components of the vector Eq. 3 become:

$$E_x = E_{Lx} + v\,B \tag{20a}$$
$$E_y = E_{Ly} - u\,B \tag{20b}$$
$$E_z = E_{Lz}. \tag{20c}$$

The third component of the plasma velocity (z-axis component, w) does not appear in the above Eqs 20a, 20b, and 20c, because even when this velocity component is not zero, it is not influenced by the magnetic field parallel to it (both being parallel to the z-axis).

Combining Eqs 20a, 20b, and 20c with Eqs 19a, 19b, and 18c gives the following 3 uncoupled explicit equations for the 3 components of the electric current density (expressed in terms of the components of the load-electrodes electric field vector, the components of the plasma velocity vector, the Hall parameter, and the electric conductivity):

$$J_x = \frac{\sigma}{1+\beta^2}(E_{Lx} + v\,B - \beta E_{Ly} + \beta u\,B) \tag{21a}$$
$$J_y = \frac{\sigma}{1+\beta^2}(E_{Ly} - u\,B + \beta E_{Lx} + \beta v\,B) \tag{21b}$$
$$J_z = \sigma E_{Lz} \tag{21c}$$

The only 2 assumptions implied in the above 3 equations are (1) neglecting the ion drift, and (2) the treatment of the magnetic field as a purely one-dimensional vector.

## 5. Reduced plasma equations for a rectangular channel with constant-voltage electrodes

### 5.1. One-dimensional uniform electric fields

In this subsection and in the rest of the analysis covered in this paper, the general mathematical description of the plasma electromagnetic behavior is customized to a special geometric case, where the electrodes are parallel horizontal surfaces, and each of them has a constant voltage. The externally applied magnetic field is still treated as one-dimensional (in the positive z-axis).

Under these conditions, the load-electrodes electric field vector becomes one-dimensional, with only one upward constant component in the positive y-axis (pointing from the lower positive cathode electrode having a higher voltage to the upper reference anode electrode having a lower voltage). Thus, this electric field vector has now only one non-zero positive component, $\vec{E_L}: (0, E_L, 0)$. Also, the source electric field vector (induced by the reaction of the moving plasma to the applied magnetic field) becomes one-dimensional, with only one downward constant component in the negative y-axis (pointing from the upper anode electrode to the lower positive cathode electrode). Thus, this electric field vector has now only one non-zero negative component, $\vec{E_S}: (0, -E_S, 0)$. Similarly, the electric field vector seen by the plasma becomes one-dimensional, with only one downward constant component in the negative y-axis. Thus, this electric field vector has now only one non-zero negative component, $\vec{E_P}: (0, -E_P, 0)$. The scalar magnitudes of these three electric fields are related as:

$$E_P = E_S - E_L. \tag{22}$$

Because all the 3 magnitude values are non-negative, the above equation implies that:

$$E_S \geq E_L \tag{23a}$$
$$E_S \geq E_P. \tag{23b}$$

With $E_{Lx} = 0$, $E_{Lz} = 0$, and $E_{Ly} = E_L$; the reduced version of Eqs. 21a, 21b, and 21c due to the one-dimensional electric fields becomes:

$$J_x = \frac{\sigma}{1+\beta^2} (v B - \beta E_L + \beta u B) \tag{24a}$$
$$J_y = \frac{\sigma}{1+\beta^2} (E_L - u B + \beta v B) \tag{24b}$$
$$J_z = 0. \tag{24c}$$

From Eq. 20a, in order to have a zero value for the x-component of both the plasma-seen electric field and the load-electrodes electric field ($E_x$ and $E_{Lx}$), the plasma must have zero velocity component in the y-axis. Thus, we must have:

$$v = 0. \tag{25}$$

Applying this condition to Eqs. 24a and 24b gives:

$$J_x = \frac{\sigma}{1+\beta^2} (-\beta E_L + \beta u B) = \frac{\sigma}{1+\beta^2} \beta (u B - E_L) \tag{26a}$$
$$J_y = \frac{\sigma}{1+\beta^2} (E_L - u B) = -\frac{\sigma}{1+\beta^2} (u B - E_L) \tag{26b}$$
$$|J_y| = \frac{\sigma}{1+\beta^2} (u B - E_L) \tag{26c}$$

where, $|J_y|$ is the magnitude of the $J_y$ current density component (which is a negative quantity). In the above equations, $J_x$ has a positive value (thus, pointing in the positive x-axis direction, which is the direction of longitudinal plasma travel). The current density component $J_y$ has a negative value (thus, pointing downward in the negative y-axis direction, from the upper anode electrode to the lower positive cathode electrode). The current density component $J_x$ is called "Hall current density." It vanishes if the Hall parameter $\beta$ is 0. This horizontal (x-component) of the current density is considered a lost uncollected flow of electrons because they do not reach the electrodes in order to pass to the external electric load. On the other hand, $J_y$ is perpendicular to the electrodes' surfaces, and it is thus the useful component of the current density vector because it is collected and utilized to power the external electric load. This component is called "Faraday current density", and it is independent of the Hall parameter. The examined design of the MHD channel in the present subsection with continuous opposite electrodes is called a "Faraday channel." In this Faraday form of the MHD generators, the useful Faraday current density is aimed to be maximized, while the loss Hall current density is aimed to be minimized. Thus, a lower Hall parameter is preferred in this design. It should be noted that Faraday channels may have slightly oblique opposite electrodes (instead of being parallel), making the channel divergent such that its cross-section area increases as the plasma moves through the channel. In supersonic plasma flows, this divergence causes the plasma to accelerate (its velocity magnitude increases).

### 5.2. Load factor

A non-dimensional scalar load factor $K_L$ can be introduced when the electric fields are uniform and one-dimensional. It is expressed as the ratio of two electric field magnitudes, $E_L$ and $E_S$:

$$K_L = E_L/E_S \tag{27}$$

thus

$$E_L = K_L E_S \tag{28}$$

where, from Eqs. 2 and 25 and with one-dimensional magnetic flux density $\vec{B}: (0, 0, B)$, we have:

$$E_S = u B \tag{29}$$

using Eq. 29 into Eq. 28 gives:

$$E_L = K_L u B. \tag{30}$$

The value of $K_L$ varies from a minimum of 0 and a maximum of 1. The minimum value of 0 indicates zero voltage difference across the external electric load, thus indicating a short-circuited load (zero-resistance load). The maximum value of 1 indicates an equal voltage difference across the external electric load and induced source voltage difference. As in electric batteries, this situation means a disconnected external electric load (open circuit condition or infinite-resistance load). Mathematically, one can write

$$0 \text{ (short circuit)} \leq K_L \leq 1 \text{ (open circuit)}. \tag{31}$$

If the plasma gas between the electrodes is treated as a lumped internal resistor with an equivalent resistance $R_P$, while the external load has a controllable resistance of $R_L$, then the load factor can be also interpreted in terms of these two resistance values as:

$$K_L = R_L/(R_L + R_P). \tag{32}$$

When the load factor is used in Eqs. 26a and 26c through replacing $E_L$ by $(K_L u B)$ as in Eq. 30, we get:

$$J_x = \frac{\sigma}{1+\beta^2} \beta u B (1 - K_L) \tag{33a}$$
$$|J_y| = \frac{\sigma}{1+\beta^2} u B (1 - K_L) \tag{33b}$$
$$J_x = \beta |J_y| \tag{33c}$$

All the individual quantities in the above Eqs. 33a, 33b, and 33c are non-negative. Eq. 33c gives another definition or interpretation for the Hall parameter, being the ratio of the Hall current density to the magnitude of the Faraday current density (when the current density is driven solely by a scalar plasma velocity and a scalar applied magnetic field).

Fig. 3 provides a sketch for the configuration of the rectangular MHD channel with constant-voltage electrodes (and a one-dimensional applied magnetic field). The electric fields are aligned with the y-axis, the plasma velocity does not have a vertical component (parallel to the y-axis), and there are both Faraday and Hall components of the current density vector. Fig. 3 also illustrates the isopotential lines (lines of constant voltage) in this configuration, which are parallel to the electrodes. If the upper anode electrode is arbitrarily assigned a reference zero voltage ($\phi_A = 0$), then the lower positive cathode electrode should have a voltage $\phi_C$ that is equal to the height h of the channel times the load-electrodes electric field. Thus:

$$\phi_C = E_L h \tag{34}$$

therefore, the constant upward electric field seen by the plasma gas is:

$$E_L = \phi_C/h \tag{35}$$

more generally, if the upper anode is assigned a non-zero voltage value, we then have:

$$E_L = (\phi_C - \phi_A)/h. \tag{36}$$

The voltage (potential) varies linearly from its minimum value at the upper reference anode electrode to its maximum value at the lower positive cathode electrode.

If the magnitude of the inclination angle of the current density vector measured from the vertical line is $\theta_J$, then:

$$\theta_J = \tan^{-1}(J_x/|J_y|) \tag{37}$$

comparing Eq. 37 to Eq. 33c, it becomes clear that the current density angle is a function of the Hall parameter.

$$\theta_J = \tan^{-1}(\beta) \tag{38}$$

For a large Hall parameter, the current density vector becomes more inclined horizontally. This is disadvantageous in the Faraday design of the MHD channel, due to the larger lost (unexploited) component of the current density vector.

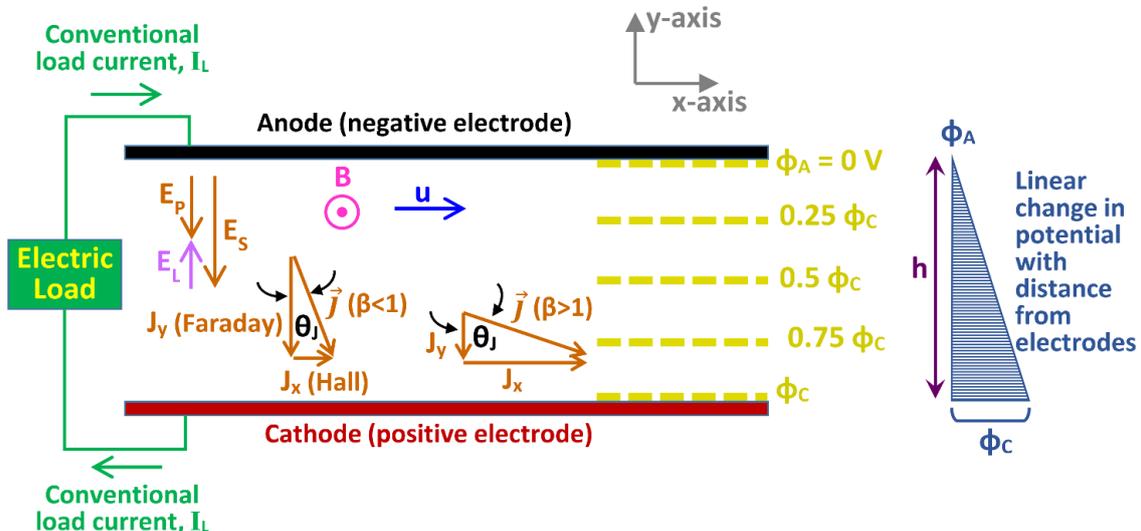

**Fig. 3:** Representation of a rectangular MHD channel with one-dimensional uniform electric fields, and a two-dimensional current density

## 6. Limiting case for Faraday channels (Zero hall parameter)

### 6.1. Implications of a zero hall parameter

From Eq. 14, which states that β = μ B, it is practically unavoidable to have a non-zero Hall parameter in a Faraday-type plasma generator. This is because in order to completely eliminate the Hall parameter (β = 0), either the applied magnetic field itself must be zero (which is a trivial condition because then no electric fields or current density fields are induced), or the electron mobility must be zero (corresponding to infinitely-many electron collisions, which is also a trivial condition because electrons should be mobile as the charge carriers). Despite this, the present subsection and the rest of the analysis covered in this paper address the hypothetical case of a Faraday MHD channel with two parallel continuous electrodes, each having a constant voltage, and with vertical current density. Thus, the Hall current density is neglected, and the conventional (positive-charge-based) current density is assumed to be totally vertical, traversing from the upper anode electrode to the lower positive cathode electrode within the plasma.

Fig. 4 is a graphical representation of this idealized case. It is similar to the configuration discussed in the previous section (5. Reduced Plasma equations for a Rectangular Channel with Constant-Voltage Electrodes), except that now the electric current density is totally vertical (totally consists of a Faraday component, no Hall component). A demonstrative example with the electrode voltages assigned arbitrary values is shown in Fig. 4, and the equipotential lines are also given, with the cathode voltage (potential) being assigned the value of ($\phi_C$ = 20 V) relative to the anode voltage. In this hypothetical example, if the spacing between the electrodes is 0.5 m, then the uniform vertical load-electrodes electric field is:

$E_L = \phi_C \div h = 20\,V \div 0.5\,m = 40\,V/m.$

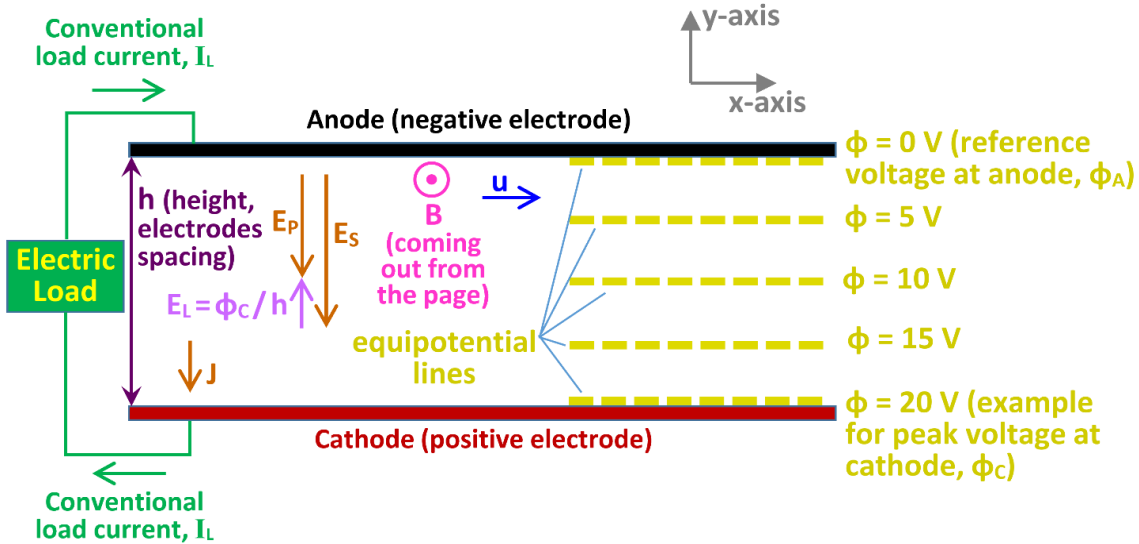

**Fig. 4:** Representation of a rectangular MHD channel with a purely-vertical current density

### 6.2. Electric current density equations

Setting β = 0 in Eqs. 33a, 33b, and 38 gives the following characteristic equations governing the current density in the present limiting case:

$J_x = 0$ (39a)
$|J_y| = \sigma\,u\,B\,(1 - K_L)$ (39b)
$\theta_J = 0$ (39c)

### 6.3. Optimum power density equation

If the plasma-facing surface area of the cathode electrode (where the conventional current density is collected from the plasma to be supplied as an electric current to an external electric load through a traditional wire) is denoted by S, and it is equal to the plasma-facing surface area of the anode electrode, then the electric current passing through the external load is:

$I_L = S\,|J_y|.$ (40)

Using Eq. 39b into Eq. 40, the load current can be related to the plasma properties (σ and u), the applied magnetic flux density B, and the load factor $K_L$ as:

$I_L = S\,\sigma\,u\,B\,(1 - K_L).$ (41)

The voltage drop $V_L$ across the external load is also the potential difference between the cathode and the anode, which can be related to the uniform load-electrodes electric field using Eq. 36, which leads to:

$V_L = \phi_C - \phi_A = E_L\,h.$ (42)

When Eq. 30, $E_L = K_L\,u\,B$, is used in the above expression, the load voltage drop can be expressed in terms of the magnitude of the plasma velocity

(scalar plasma speed), the magnitude of the applied magnetic flux density, and the load factor as:

$$V_L = k_L u\, B\, h \quad (43)$$

the above equation can be rewritten as:

$$V_L = k_L V_{OC} \quad (44a)$$
$$V_{OC} = u\, B\, h \quad (44b)$$

where, $V_{OC}$ is the open-circuit voltage, or the maximum (without any load connected) voltage difference between the two electrodes (the positive-voltage cathode and the reference-voltage anode) that the plasma generator can provide as a battery-like voltage source. As in traditional electric circuits analysis, the DC power ($P_L$) delivered to the load is the product of the electric current passing through it and the voltage drop across it. Thus:

$$P_L = V_L I_L \quad (45)$$

using Eqs. 41 and 43 in the above equation gives:

$$P_L = [k_L u\, B\, h]\,[S\, \sigma\, u\, B\, (1 - K_L)] = (h\, S) \sigma\, u^2\, B^2\, [K_L (1 - K_L)]. \quad (46)$$

However, the product (h S) is the volume ($\tilde{V}$) of the plasma gas inside the Faraday MHD channel (the space confined between the upper and lower electrodes and the front and back electrically-insulated walls). Thus:

$$P_L = \tilde{V}\, \sigma\, u^2\, B^2\, [K_L (1 - K_L)] \quad (47)$$

The above electric power output expression in Eq. 47 reveals a quadratic dependence of the electric power output on the plasma speed as well as on the applied magnetic flux density. Thus, these two quantities have a more important role in the extracted electric power compared to the plasma electric conductivity and the channel volume, which affect the electric output power linearly.

The electric power output per unit volume of plasma (the volumetric power density or $P_{LV}$), is obtained after dividing the electric DC power delivered to the load by the channel volume. This gives:

$$P_{LV} = \sigma\, u^2\, B^2\, [K_L (1 - K_L)]. \quad (48)$$

If the baseline quantity ($\sigma\, u^2\, B^2$) in Eq. 48 is treated as a constant within the MHD channel, then the volumetric power density becomes a function of the load factor alone, through the multiplier: $K_L (1 - K_L)$. Fig. 5 shows how this multiplier varies as the load factor $K_L$ changes from its minimum ($K_L = 0$, shorted load) to its maximum ($K_L = 1$, disconnected load). It can also be shown mathematically or graphically that this multiplier has a single maximum at $K_L = 0.5$, at which it takes the peak value of 0.25. Thus, the electric DC power to the load is maximized when the load resistance is matched with the equivalent internal resistance of the plasma gas. This means that the induced electric voltage magnitude (u B h) is equally distributed as an internal voltage drop (0.5 u B h) across the plasma and an external voltage drop (0.5 u B h) across the load. This particular load value (the matched load value) is the optimum for the best utilization of the plasma generator as a powering source.

Therefore, the maximum matched-load power output from the Faraday-type rectangular MHD channel having constant properties (electric conductivity, plasma speed, and magnetic flux density) with an idealized operation by neglecting the Hall parameter (assumed to be zero) is:

$$P_{LV,opt} = 0.25\, \sigma\, u^2\, B^2 \quad (49)$$

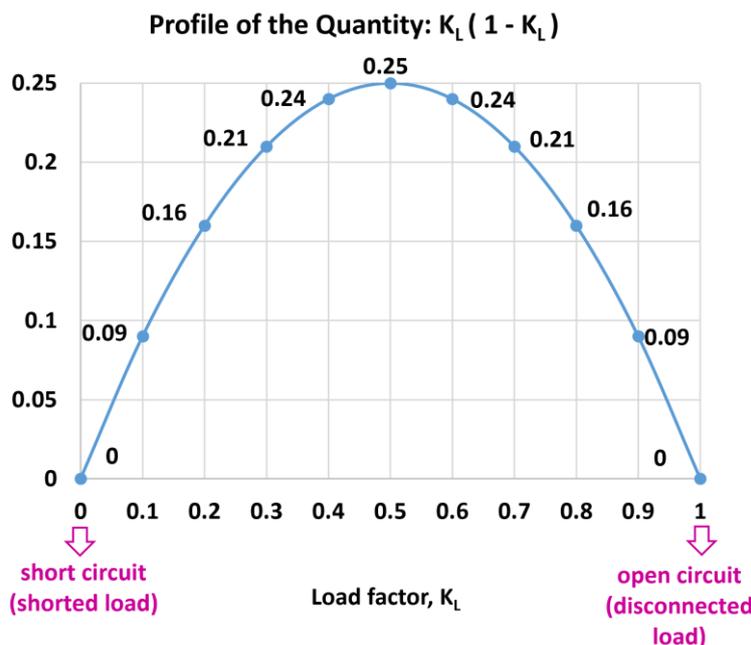

**Fig. 5:** Variation of $[K_L (1 - K_L)]$ with the load factor $K_L$, for the entire range of $K_L$ (from 0 to 1)

## 6.4. Estimated ideal volumetric power density

The final part of this study in the present subsection is making a suitable numerical estimation of the above-described optimum volumetric power density for a rectangular constant-voltage Faraday MHD channel. This requires selecting typical values for the 3 quantities that appear on the right-hand side of Eq. 49; namely σ, u, and B.

For the plasma electric conductivity σ, a value of 20 S/m is selected here (Sawhney and Verma, 1990; Mikheev et al., 2006; Pavshuk and Panchenko, 2008). It should be noted that this particular plasma property can vary largely depending on the temperature, seed type, and seed amount.

For the plasma streamwise speed u, a broad range is possible. A mild supersonic speed is considered here. Subsonic speeds are also permitted (Mcclaine et al., 1985; Aithal, 2009), but the higher supersonic speeds allow much more electric outputs. It is common to relate such a high gas speed to the local seed of sound through the Mach number M, as (NTRS, 1953):

$$M = u/a \qquad (50)$$

Supersonic fluid speeds have $M > 1$, while subsonic fluid speeds have $M < 1$. The speed of sound can be computed from the absolute temperature T, the specific gas constant R, and the specific heat ratio γ as (Leishman, 2023):

$$a = \sqrt{\gamma R T} \qquad (51)$$

While air and other diatomic gases have a specific heat ratio near 1.4 at room temperatures (Seibert and Nieh, 2016), this quantity drops for triatomic gases (Jost, 1996) and also drops at elevated temperatures (Ceviz and Kaymaz, 2005). Assuming a plasma temperature of 2800 K that consists mainly of carbon dioxide and hot water vapor (steam), a good specific heat ratio would then be approximately 1.17 (Marzouk, 2018b).

The specific gas constant for carbon dioxide is 188.9 J/kg.K, while the specific gas constant for water vapor is 461.5 J/kg.K (Wong and Segall, 2020). Assuming a plasma gas consisting of equal mole fractions (equal volumes) of these two common combustion-product gases (the seed is neglected because it is a minor component), it can be shown that the specific gas constant R of that selected mixture is the harmonic mean of the individual specific gas constants, $R_{CO2}$ and $R_{H2O}$ (Kuo, 2005). This is computed as:

$$R = 2 R_{CO2} R_{H2O} / (R_{CO2} + R_{H2O}) \qquad (52)$$

which gives R = 268.1 J/kg.K. Therefore, Eq. 51 gives an estimated sonic speed of 937.2 m/s within the plasma at 2800 K. Selecting a suitable plasma Mach number near 2.134 (Aithal, 2008; Pengyu et al., 2016), the estimated supersonic plasma speed becomes 2000 m/s.

For the magnetic flux density B, a field strength of 5 T is selected here (Murray et al., 2003; Macheret et al., 2004). With this, Eq. 49 gives an estimated optimum ideal volumetric power output of $P_{LV}=0.5×10^9$ W/m$^3$ (0.5 GW/m$^3$ or 500 MW/m$^3$). This is a very high power density compared to conventional heat engines (Khazin et al., 2019), reflecting extremely intense utilization of volume. For example, if a target power output of 1 GW (1000 MW) is sought, which is typical of commercial power plants, then a channel volume of only $\tilde{V}_{1GW}=2$ m$^3$ is needed. In a previous study (Marzouk, 2017), a volumetric power density was estimated as 194 MW/m$^3$, but with an estimated plasma speed of 1200 m/s, a plasma conductivity of 15 S/m, and a magnetic field strength of 6 T. That benchmarking case is consistent with the analysis provided here, where Eq. 49 gives an estimated volumetric power density of 194.4 MW/m$^3$ with these published parameters. In another preliminary experimental investigation (Li et al., 2011), a much smaller volumetric power density of 4.7971 MW/m$^3$ was reported. That estimate corresponds to a plasma speed of 1959 m/s (close to the value adopted here), and an electric conductivity of 20 S/m (identical to the value adopted here). However, the magnetic flux density was only 0.5 T. With these published conditions, Eq. 49 gives a power density value of 4.7971 MW/m$^3$, which is the same as the reported value, and this additional benchmarking case further supports the validity of the analysis presented here. Table 1 summarizes the parameters adopted in the preceding power estimation.

**Table 1:** Power output estimation for an idealized MHD channel (Faraday type) with constant properties

| Quantity | Selected or computed | Value | Unit |
| --- | --- | --- | --- |
| σ | Selected | 20 | S/m |
| T | Selected | 2800 | K |
| R | Selected | 268.1 | J/kg.K |
| γ | Selected | 1.17 | - |
| a | Computed | 937.2 | m/s |
| M | Selected | 2.134 | - |
| u | Computed | 2000 | m/s |
| B | Selected | 5 | T |
| $P_{LV}$ | Computed | 500 | MW/m$^3$ |
| $\tilde{V}_{1GW}$ | Computed | 2 | m$^3$ |

## 7. Conclusion

In this paper, the mathematical description was presented for the electric current density field and the electric field within a moving plasma (seeded ionized gas) inside a channel for the purpose of electric power extraction, due to the reaction of the electrically conducting ionized gas to an applied external magnetic field. The modeling was presented first for a general fully three-dimensional problem that is not tied to a particular channel geometry, and then for a still-general geometry-independent problem but under partial three-dimensionality after restricting the magnetic field to be one-dimensional. The MHD (magnetohydrodynamics) plasma problem modeling was then simplified more to the case where the fields are one-dimensional except for the electric current density, under the geometric condition of parallel electrodes and the electric condition of constant-voltage electrodes. Finally, the limiting case of the zero Hall parameter (where all fields can be represented by constant scalars) was investigated, and the expression for the optimum volumetric electric power density was given. A quantitative estimate for this quantity was given and discussed.

## List of symbols

| | |
|---|---|
| $a$ | Speed of sound- SI unit: m/s |
| $B$ | Magnitude of the external magnetic flux density field applied on plasma- SI unit: T (tesla) |
| $\vec{B}$ | Vector form of B- SI unit: T (tesla) |
| $\vec{d}$ | Mean drift velocity vector of electrons (average electrons velocity vector relative to their carrying medium)- SI unit: m/s |
| $E_L$ | Magnitude of the applied electric field on the plasma due to the voltage difference at the electrodes surrounding it. These electrodes are possibly connected to an external electric load.- SI unit: V/m |
| $\vec{E_L}$ | Vector form of $E_L$- SI unit: V/m |
| $E_P$ | Magnitude of the consumed electric field within the plasma due to its internal resistance for passing current density- SI unit: V/m |
| $\vec{E_P}$ | Vector form of $E_P$- SI unit: V/m |
| $E_S$ | Magnitude of the source (induced) electric field consumed due to the motion of the plasma within an applied magnetic field- SI unit: V/m |
| $\vec{E_S}$ | Vector form of $E_S$- SI unit: V/m |
| $e$ | Magnitude of the electron charge- SI unit: e = 1.602176634 × 10⁻¹⁹ C (coulomb) (Gilbey, 2023) |
| eV | Electronvolt or electron-volt, a small unit of energy representing the required energy for moving a charge equal to 1 e across a potential difference of 1 V - SI unit: eV = 1.602176634 × 10⁻¹⁹ J (joule) (Di Sia, 2021) |
| $J$ | Magnitude of the current density in the plasma when it is assumed to be one-dimensional (in the negative y-axis)- SI unit: A/m² |
| $\vec{j}$ | Vector form of current density within the plasma- SI unit: A/m² |
| $J_x$ | Horizontal component (x-axis) of the current density vector within the plasma- SI unit: A/m² |
| $J_y$ | Vertical component (y-axis) of the current density vector within the plasma- SI unit: A/m² |
| $J_z$ | Outward component (z-axis) of the current density vector within the plasma- SI unit: A/m² |
| $h$ | Height of the plasma channel (distance between the electrodes)- SI unit: m |
| $I_L$ | Electric current passing through the external electric load, assigned a conventional direction (from a positive terminal to a negative terminal). Thus, it is considered to flow from the positive cathode electrode to the negative/reference anode electrode through a solid conductor outside the plasma channel.- SI unit: A (ampere) |
| $K_L$ | Load factor, which is the ratio of load resistance across the external load to the sum of this external load resistance and the lumped internal resistance of the plasma. It is also the ratio of the magnitude of the applied electric field to the magnitude of the source electric field.- SI unit: no unit |
| M | Mach number - SI unit: no unit |
| m | Mass of electron- SI unit: m = 9.1093837 × 10⁻³¹ kg (Miyashita, 2023) |
| $P_L$ | Electric power delivered to the external electric load, which is a direct-current (DC) electric power consumption - SI unit: W (watt) |
| $P_{LV}$ | Volumetric power density, which is the electric power output to the load per unit volume of plasma, $P_{LV} = P_L/\tilde{V}$- SI unit: W/m³ |
| $P_{LV,opt}$ | Optimum volumetric power density (for a matched load)- SI unit: W/m³ |
| R | Specific gas constant- SI unit: J/kg.K |
| S | Surface area of one face of the electrode plate- SI unit: m² |
| T | Absolute temperature- SI unit: K (kelvin) |
| u | Horizontal component (x-axis) of the velocity vector of the plasma gas- SI unit: m/s |
| $\vec{u}$ | Vector form of the velocity of the plasma- SI unit: m/s |
| $\tilde{V}$ | Volume of the plasma channel- SI unit: m³ |
| $\tilde{V}_{1GW}$ | Volume of the plasma channel for 1 GW electric power output - SI unit: m³ |
| $V_{OC}$ | Open-circuit voltage- SI unit: V (volt) |
| $V_L$ | Voltage drop across the external electric load SI unit: V (volt) |
| v | Vertical component (y-axis) of the velocity vector of the plasma gas- SI unit: m/s |
| w | Outward component (z-axis) of the velocity vector of the plasma gas- SI unit: m/s |
| β | Hall parameter, which is the ratio of the mean free path of electrons within the plasma to the radius of the circular motion of a moving electron subject to a magnetic field. It is also the product of the mean cyclotron angular frequency of electron and the mean free time between collisions of electron. It is also the ratio of the horizontal x-axis component of the current density vector (the Hall component) to the magnitude of the vertical y-axis component of the current density vector (the Faraday component).- SI unit: no unit |
| γ | Specific heat ratio (ratio of the specific heat at constant pressure to the specific heat at constant volume) - SI unit: no unit |
| ϕ | Potential (voltage); it is constant at each "virtual" isopotential line within the plasma- SI unit: V (volt) |
| $ϕ_A$ | Potential (voltage) at the anode electrode (the negative/reference electrode in the case of MHD generators)- SI unit: V (volt) |
| $ϕ_C$ | Potential (voltage) at the cathode electrode (the positive electrode in the case of MHD generators)- SI unit: V (volt) |
| μ | Electron mobility, which is the drift speed gained by an electron when it is placed in a unit electric field, $μ = eτ/m$- SI unit: (m/s)/(V/m) or m²/s.V |
| $θ_J$ | Inclination angle of the electric current density vector, measured from the vertical (y-axis)- SI unit: radian |
| σ | Electric conductivity of the plasma gas- SI unit: S/m (siemens/m) or A/m.V |
| τ | Mean free time between collisions for an electron- SI unit: s (second) |

## Compliance with ethical standards

## Conflict of interest

The author(s) declared no potential conflicts of interest with respect to the research, authorship, and/or publication of this article.

# References

Aithal SM (2008). Analysis of optimum power extraction in a MHD generator with spatially varying electrical conductivity. International Journal of Thermal Sciences, 47(8): 1107-1112. https://doi.org/10.1016/j.ijthermalsci.2007.09.001

Aithal SM (2009). Characteristics of optimum power extraction in a MHD generator with subsonic and supersonic inlets. Energy Conversion and Management, 50(3): 765-771. https://doi.org/10.1016/j.enconman.2008.09.037

Aminudin MA, Kamarudin SK, Lim BH, Majilan EH, Masdar MS, and Shaari N (2023). An overview: Current progress on hydrogen fuel cell vehicles. International Journal of Hydrogen Energy, 48(11): 4371-4388. https://doi.org/10.1016/j.ijhydene.2022.10.156

Bera TK, Bohre AK, Ahmed I, Bhattacharya A, and Bhowmik PS (2022). Magnetohydrodynamic (MHD) power generation systems. In: Bohre AK, Chaturvedi P, Kolhe ML, and Singh SN (Eds.), Planning of hybrid renewable energy systems, electric vehicles and microgrid: Modeling, control and optimization: 905-929. Springer, Singapore, Singapore. https://doi.org/10.1007/978-981-19-0979-5_34

Blackman VH, Jones Jr MS, and Demetriades A (1961). MHD power generation studies in rectangular channels. In the Proceedings of the 2nd Symposium on the Engineering Aspects of Magnetohydrodynamics, Columbia University Press, Philadelphia, USA: 180–210.

Borghi CA, Massarini A, Mazzanti G, and Ribani PL (1992). Steady state descriptions of MHD plasma flows. In the 11th International Conference on MHD Electrical Power Generation, International Academic Publishers, 3: 770-775.

Cébron D, Viroulet S, Vidal J, Masson JP, Viroulet P (2017). Experimental and theoretical study of magnetohydrodynamic ship models. PLOS ONE 12(10): e0186166. https://doi.org/10.1371/journal.pone.0186166 PMid:28977028 PMCid:PMC5627937

Ceviz MA and Kaymaz İ (2005). Temperature and air–fuel ratio dependent specific heat ratio functions for lean burned and unburned mixture. Energy Conversion and Management, 46(15–16): 2387-2404. https://doi.org/10.1016/j.enconman.2004.12.009

Chen XQ, Zhao LZ, and Peng AW (2022). Performance analysis of a new helical channel seawater MHD thruster. In the 32nd International Ocean and Polar Engineering Conference, Shanghai, China: ISOPE-I-22-526.

Cheng K, Wang Y, Xu J, Qin J, and Jing W (2022). A novel liquid metal MHD enhanced Closed-Brayton-Cycle power generation system for hypersonic vehicles: Thermodynamic analysis and performance evaluation with finite cold source. Energy Conversion and Management, 268: 116068. https://doi.org/10.1016/j.enconman.2022.116068

Danilevich JB, Antipov VN, Kruchinina IY, Khozikov YP, and Ivanova AV (2010). Prospective permanent magnet turbogenerator design for local power engineering. In the XIX International Conference on Electrical Machines, IEEE, Rome, Italy: 1-4. https://doi.org/10.1109/ICELMACH.2010.5608254

Davidson PA (2001). An introduction to magnetohydrodynamics. Cambridge University Press, Cambridge, USA. https://doi.org/10.1017/CBO9780511626333

Di Sia P (2021). Birth, development and applications of quantum physics: a transdisciplinary approach. World Scientific News, 160: 232–246. https://doi.org/10.31219/osf.io/2tyxe

Esmaeilzadehazimi MA, Manesh MHK, Heleyleh BB, and Modabbaer HV (2019). 4E analysis of integrated MHD-combined cycle. International Journal of Thermodynamics, 22(4): 219-228. https://doi.org/10.5541/ijot.570540

Everett CN and Ryan CN (2023). A linear magnetic reconnection based plasma thruster for spacecraft propulsion. American Institute of Aeronautics and Astronautics SciTech 2023 Forum. https://doi.org/10.2514/6.2023-0448.vid

Gilbey JD (2023). SI units. Anaesthesia and Intensive Care Medicine, 24(4): 244–247. https://doi.org/10.1016/j.mpaic.2022.12.030

Ishikwa M, Yuhara M, and Fujino T (2007). Three-dimensional computation of magnetohydrodynamics in a weakly ionized plasma with strong MHD interaction. Journal of Materials Processing Technology, 181(1–3): 254-259. https://doi.org/10.1016/j.jmatprotec.2006.03.032

Jones AR (1985). MHD advanced power train: Phase 1, final report: Volume 2, development program plan (No. DOE/PC/60575-T1-Vol. 2). Westinghouse Electric Corp, Advanced Energy Systems Division, Pittsburgh, USA. https://doi.org/10.2172/93533

Jost R (1996). The cooling of internal degrees of freedom of polyatomic molecules in supersonic free jets. In: Fausto R (Ed.), Low temperature molecular spectroscopy: 249-270. Springer Netherlands, Dordrecht, Netherlands. https://doi.org/10.1007/978-94-009-0281-7_10

Kayukawa N (2004). Open-cycle magnetohydrodynamic electrical power generation: A review and future perspectives. Progress in Energy and Combustion Science, 30(1): 33-60. https://doi.org/10.1016/j.pecs.2003.08.003

Khazin LM, Tarasov PI, and Furzikov VV (2019). Use of gas-turbine engines for mining dump trucks in the conditions of the north. Perm Journal of Petroleum and Mining Engineering, 29(3): 290-300. https://doi.org/10.15593/2224-9923/2019.3.8

Kuo KK (2005). Principles of combustion. 2nd Edition, John Wiley and Sons, Hoboken, USA.

Leishman JG (2023). Introduction to aerospace flight vehicles. Embry-Riddle Aeronautical University, Daytona Beach, USA. https://doi.org/10.15394/eaglepub.2022.1066

Li Y, Li Y, Lu H, Zhu T, Zhang B, Chen F, and Zhao X (2011). Preliminary experimental investigation on MHD power generation using seeded supersonic argon flow as working fluid. Chinese Journal of Aeronautics, 24(6): 701-708. https://doi.org/10.1016/S1000-9361(11)60082-4

Liu Z, Tian B, Li Y, Guo Z, Zhang Z, Luo Z, Zhao L, Lin Q, Lee C, and Jiang Z (2023). Evolution of thermoelectric generators: From application to hybridization. Small, 19(48): 2304599. https://doi.org/10.1002/smll.202304599 PMid:37544920

Macheret SS, Schneider M, Murray R, Zaidi S, Vasilyak L, and Miles R (2004). RDHWT/MARIAH II MHD modeling and experiments review. In the 24th AIAA Aerodynamic Measurement Technology and Ground Testing Conference, Portland, USA. https://doi.org/10.2514/6.2004-2485

Markoulakis E, Vanderelli T, Frantzeskakis L (2022). Real time display with the ferrolens of homogeneous magnetic fields. Journal of Magnetism and Magnetic Materials, 541: 168576. https://doi.org/10.1016/j.jmmm.2021.168576

Marzouk OA (2017). Combined oxy-fuel magnetohydrodynamic power cycle. ArXiv Preprint ArXiv:1802.02039. https://doi.org/10.48550/arXiv.1802.02039

Marzouk OA (2018a). Multi-physics mathematical model for weakly-ionized plasma flows. American Journal of Modern Physics, 7(2): 87-102. https://doi.org/10.11648/j.ajmp.20180702.14

Marzouk OA (2018b). Assessment of three databases for the NASA seven-coefficient polynomial fits for calculating thermodynamic properties of individual species. International Journal of Aeronautical Science and Aerospace Research, 5(1): 150-163. https://doi.org/10.19070/2470-4415-1800018

Marzouk OA (2022). Land-Use competitiveness of photovoltaic and concentrated solar power technologies near the Tropic of Cancer. Solar Energy, 243: 103-119. https://doi.org/10.1016/j.solener.2022.07.051

Marzouk OA (2023). Cantera-based Python computer program for solving steam power cycles with superheating. International Journal of Emerging Technology and Advanced Engineering, 13(3): 63-73. https://doi.org/10.46338/ijetae0323_06

Mcclaine A, Swallom D, and Kessler R (1985). Experimental investigation of subsonic combustion driven MHD generator performance. Journal of Propulsion and Power, 1(4): 263-269. https://doi.org/10.2514/3.22792

Messerle HK (1995). Magnetohydrodynamic electrical power generation. John Wiley and Sons, Hoboken, USA.

Mikheev AV, Kayukawa N, Okinaka N, Kamada Y, and Yatsu S (1991). High-temperature coal-syngas plasma characteristics for advanced MHD power generation. IEEE Transactions on Energy Conversion, 21(1): 242-249. https://doi.org/10.1109/TEC.2005.847994

Miyashita T (2023). A fine-structure constant can be explained using the electrochemical method. Journal of Modern Physics, 14(2): 160-170. https://doi.org/10.4236/jmp.2023.142011

Murray R, Zaidi S, Kline J, Shneider M, Macheret S, and Miles R (2003). Investigation of a Mach 3 cold air MHD channel. In the 34th AIAA Plasmadynamics and Lasers Conference. Orlando, Florida. https://doi.org/10.2514/6.2003-4282

NCBI (2023). Ionization energy in the periodic table of elements. National Center for Biotechnology Information, Bethesda, USA.

NTRS (1953). Equations, tables, and charts for compressible flow. NASA Technical Reports Server, Ames Aeronautical Laboratory, Ames Aeronautical Staff, Moffett Field, USA.

Okamura T, Kabashima S, Shioda S, and Sanada Y (1985). Superconducting magnet for a disc generator of the FUJI-1 MHD facility. Cryogenics, 25(9): 483-491. https://doi.org/10.1016/0011-2275(85)90069-4

Okunoa Y, Okamura T, Yoshikawa K, Suekane T, Tsuji K, Okubo M, Maeda T, Murakami T, Yamasaki H, Kabashima S, Shioda S, and Hasegawa Y (1999). High enthalpy extraction experiments with Fuji-1 MHD blow-down facility. Energy Conversion and Management, 40(11): 1177-1190. https://doi.org/10.1016/S0196-8904(99)00006-0

Panchenko VP (2002). Preliminary analysis of the "Sakhalin" world largest pulsed MHD generator. In the 33rd Plasmadynamics and Lasers Conference, Maui, Hawaii. https://doi.org/10.2514/6.2002-2147

Pavshuk VA and Panchenko VP (2008). Open-cycle multi-megawatt MHD space nuclear power facility. Atomic Energy, 105: 175-186. https://doi.org/10.1007/s10512-008-9082-1

Pengyu Y, Bailing Z, Yiwen L, Yutian W, Chengduo D, Hao F, Ling G (2016). Investigation of MHD power generation with supersonic non-equilibrium RF discharge. Chinese Journal of Aeronautics, 29(4): 855-862. https://doi.org/10.1016/j.cja.2016.06.018

Raja AK, Srivastava AP, and Dwivedi M (2006). Power plant engineering. New Age International, New Delhi, India.

Rosa RJ, Krueger CH, and Shioda CH (1991). Plasmas in MHD power generation. IEEE Transactions on Plasma Science, 19(6): 1180-1190. https://doi.org/10.1109/27.125040

Sarkar DK (2017). Chapter 1: General description of thermal power plants. In: Sakar D (Ed.), Thermal power plant: Pre-operational activities: 1–31. Elsevier, Amsterdam, Netherlands. https://doi.org/10.1016/B978-0-08-101112-6.00001-0

Sawhney BK and Verma SS (1990). International journal of energy research. Atomic Energy, 14(4): 433-447. https://doi.org/10.1002/er.4440140408

Seibert ML and Nieh S (2016). Control of an air siphon nozzle using hydrogen and gases other than air. International Journal of Hydrogen Energy, 41(5): 683-689. https://doi.org/10.1016/j.ijhydene.2015.10.068

Sen D, Tunç KM, and Günay ME (2021). Forecasting electricity consumption of OECD countries: A global machine learning modeling approach. Utilities Policy, 70: 101222. https://doi.org/10.1016/j.jup.2021.101222

Smit BA, Zyl EV, Joubert JJ, Meyer W, Prévéral S, Lefèvre CT, and Venter SN (2018). Magnetotactic bacteria used to generate electricity based on Faraday's law of electromagnetic induction. Letters in Applied Microbiology, 66(5): 362-367. https://doi.org/10.1111/lam.12862 **PMid:29432641**

Takeishi K and Krewinkel R (2023). Advanced gas turbine cooling for the carbon-neutral era. International Journal of Turbomachinery, Propulsion and Power, 8(3): 19. https://doi.org/10.3390/ijtpp8030019

Wong YQ and Segall P (2020). Joint inversions of ground deformation, extrusion flux, and gas emissions using physics-based models for the Mount St. Helens 2004–2008 eruption. Geochemistry, Geophysics, Geosystems, 21(12): e2020GC009343. https://doi.org/10.1029/2020GC009343

Woodside CR, Richards G, Huckaby ED, Marzouk OA, Haworth DC, Celik IB, Ochs T, Oryshchyn D, Strakey PA, Casleton KH, Pepper J, Escobar-Vargas J, and Zhao X (2012). Direct power extraction with oxy-combustion: An overview of magnetohydrodynamic research activities at the NETL-regional university alliance (RUA). In the 29th Annual International Pittsburgh Coal Conference, Pittsburgh, USA: 825-843.

Wright JK (1963). Paper 3: Paper 4: Physical principles of MHD generation. Proceedings of the Institution of Mechanical Engineers, Conference Proceedings, 178(8): 39-47. https://doi.org/10.1243/PIME_CONF_1963_178_195_02

Zhang F, Wang B, Gong Z, Zhang X, Qin Z, and Jiao K (2023). Development of photovoltaic-electrolyzer-fuel cell system for hydrogen production and power generation. Energy, 263(A): 125566. https://doi.org/10.1016/j.energy.2022.125566

Zhao B, Li Y, Zhou C, and Zheng J (2022). Simulation of magnetoplasmadynamic thruster with a fluid model. In 3rd International Conference on Artificial Intelligence and Electromechanical Automation, Changsha, China, 12329: 123290Z. https://doi.org/10.1117/12.2646863

Zhou J, Wei T, and An X (2023). Combining non-thermal plasma technology with photocatalysis: A critical review. Physical Chemistry Chemical Physics, 25: 1538-1545. https://doi.org/10.1039/D2CP04836A **PMid:36541425**